# Effect of heat treatment on the structural and microstructural properties of the $Co_2$-Y hexaferrites

Osama Mohsen[1], Sami H. Mahmood[1*], Ahmad Awadallah[1], Yazan Maswadeh[2]

[1]Physics Department, The University of Jordan

[2]Physics Department, Central Michigan University

**Abstract:** The effect of heat treatment on the structural and micro-structural properties of the $Co_2$-Y ($Ba_2Co_2Fe_{12}O_{22}$) was investigated by means of X-ray diffraction and the Rietveld refinement. The samples were synthesized using high-energy ball milling technique. Phase identification and Rietveld refinement of the samples confirmed the presence of a single Y-type phase that is consistent with reported patterns. Analysis of the lattice constants obtained using the Rietveld refinement confirmed the presence of inner distortion, which was responsible for increasing the lattice constants. However, after the heat treatment, the lattice constants were in good agreement with the reported pattern. The structural analysis using quadratic elongation revealed distortions in the crystal structure. The structural analysis revealed some differences in the cation-anion distances in some sites, while in other sites, these distances remained the same. The use of the Rietveld refinement to obtain micro-structural information about the size and the strain was reported. Heat treatment induced diffusion between crystal domains leading to an increase in crystallite size.

# Introduction

Because of their importance for a wide range of applications, cost and easy low manufacturing, as well as the ability to fine-tune their magnetic and electrical properties, hexagonal ferrites have attracted the attention of engineers and scientists since their discovery in the 1950s [1, 2]. Their favorable properties for permanent magnets, high density magnetic recording, ultrahigh frequency microwave devices and microwave absorption, placed hexagonal ferrites at the top of the list of materials in the market today. The properties of the ferrites were modified by various techniques, the most efficient of which is metal substitutions and appropriate heat treatment [1, 3-15].

The crystal structure of the basic type of hexagonal ferrites, the M-type hexagonal ferrite ($BaFe_{12}O_{19}$), was investigated by Philips Laboratories after World War (II). Continued research on ferrites lead to the discovery of more complicated forms of hexaferrites, namely, the Y, W, Z, X, and U-type hexagonal ferrites [1, 2, 16].

The unit cell of the Y-type hexagonal ferrite ($Ba_2Me_2Fe_{12}O_{22}$) is built from sequential stacking of the S and the T-blocks in a sequence of (TST'ST"S") where the primes indicate a rotation about the $c$-axis with 120 degrees [1, 8, 17, 18]. Each unit cell contains three molecules, one in each TS block stacking. The length of the unit cell along the $c$-axis is 43.56 Å, while the length along the $a$-axis is 5.88 Å.

Y-type hexaferrite crystallizes in a structural type characterized by the space group of R-3m. The Y-type has 2 divalent (Me) cations per formula unit, and all the cations ($Me^{+2}$ and $Fe^{+3}$) are distributed over six crystallographic sites (table 1). The type of these divalent cations, as well as the site they occupy within the unit cell can result in significant modifications of the structural and magnetic properties for this type of ferrite. [1, 13, 19-24]. However, detailed structural and microstructural analyses of Y-type hexaferrites was found limited in the available literature.

Table 1: Crystallographic positions occupied by various cations in the Y-type structure

| Block | Coordination | Site |
|---|---|---|
| S | Tetrahedral | $6c_{IV}$ |
| S | Octahedral | $3a_{VI}$ |
| T | Octahedral | $6c_{VI}$ |
| T | Tetrahedral | $6c_{IV}$* |
| T | Octahedral | $3b_{VI}$ |
| T-S | Octahedral | $18h_{VI}$ |

The present work is concerned with the fabrication and investigation of the structural and microstructural characteristics of the $Co_2$-Y compound. The effect of heat treatment on the structural and microstructural properties is addressed.

**Experimental**

The precursors of $Ba_2Co_2Fe_{12}O_{22}$ ($Co_2$-Y) sample were synthesized using the wet high-energy ball milling method. The starting powders were stoichiometric ratios of high purity (> 98%) $BaCO_3$, $Fe_2O_3$ and Co. The resulting mixtures were milled for 16 hours, in intervals of 10 minutes each, separated by intervals of 5 minutes to avoid overheating during the process. The process was performed using Fritsch Pulverissette-7 ball mill. The resulting muddy mass was left to dry at room temperature, and then the dry powder was collected.

About 0.8 g of the powder was pressed into a disk (~1.5 cm diameter) under a force of 4-5 tons. The disk was then sintered for 2 h at 1200° C, with a heating rate of 10° C per minute, in a zirconium oxide crucible, and then left to cool gradually down to room temperature. The effect of heat treatment on the structural and microstructural properties was investigated by carrying out additional measurements on a sample sintered at 1200° C and subsequently annealed at 600° C for 4 hours ($Co_2$-$Y_R$).

The XRD sample was prepared by grinding parts of the disk for about 30 minutes using a mortar and pestle to obtain fine and smooth powder, and avoid any preferred orientation. Double scotch tape was then placed on a clean stainless steel plate, and a flat layer of the sample was prepared by sprinkling the powder on the sticky face of the tape. X-ray diffraction measurement on the sample was performed using 7000-Shimadzu diffractometer with Cu-K$_\alpha$ radiation ($\lambda_{\alpha1}$ = 1.540560 Å, $\lambda_{\alpha2}$ = 1.544390Å). The angular range was set to be 2θ = 20° -70° with a step-size of 0.01° and a scan speed of 0.5° per minute.

For phase identification, PDF-2(2003) database was used. The whole pattern analysis, the standard Rietveld refinement and the micro-structural analysis, was performed using FullProf software package (Fullprof Suite (2.05)) [25, 26].

## Results and Discussions

X-ray diffraction patterns of Co$_2$-Y and Co$_2$-Y$_R$ samples (figure 1) indicated that each sample was a single-phase consistent with the Ba$_2$Co$_2$Fe$_{12}$O$_{22}$ Y-type hexaferrite standard (JCPDS 00-044-0206) with no secondary phases. The pattern for Co$_2$-Y$_R$ sample showed some degree of preferred orientation along the *c*-axis direction as indicated by the intensity of the (1 0 13) reflection located at about 32°.

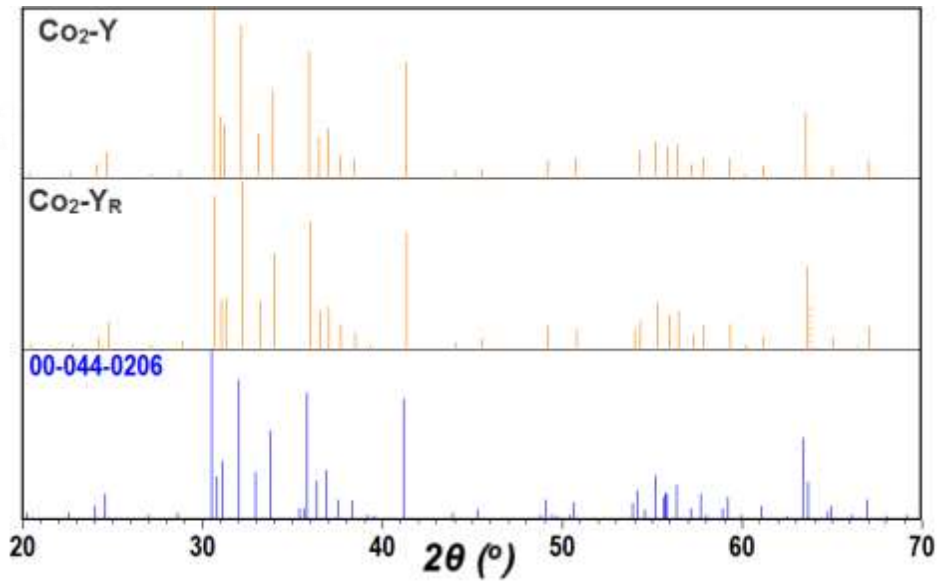

Fig. 1: X-ray diffraction patterns of samples Co$_2$-Y and Co$_2$-Y$_R$

The initial parameters for the refinement were obtained from the neutron diffraction data of Collomb et al. [20]. The adoption of these parameters for our analysis is based on the findings of a recent XRD study [8], which confirmed that these parameters did not differ noticeably from those obtained from the neutron diffraction data. It was reported that $Co^{+2}$ cations can occupy 4 different sites, one tetrahedral ($6c_{IV}$), and 3 octahedral sites (3b, 18h and 3a) [20]. Since the diffracted intensity is similar for scattering by either $Fe^{+3}$ or $Co^{+2}$, the occupancy ratio consistent with the stoichiometric concentrations of the two cations was adopted from a previous study [27].

The experimental data and the results of the Rietveld refinement for both samples ($Co_2$-Y and $Co_2$-$Y_R$) are shown in Fig. 2. Except for some differences in the residual (the difference between the experimental data and the refined pattern) in the angular range $2\theta = 20° - 40°$, where major peaks overlap, the residual in the patterns is negligibly small.

The result of the refinement and the lattice constants for the sample $Co_2$-Y and $Co_2$-$Y_R$ are listed in table (2). $\chi^2$ values, as well as the other reliability factors ($R_F$ and $R_B$) are relatively small, indicating the reliability of the fit. The lattice constants for the sample annealed at 600° C appear to decrease slightly, approaching the values for the standard pattern ($a = b = 5.86$ Å, $c = 43.50$ Å). The slightly higher lattice constants of the sample ($Co_2$- Y) are attributed to inner distortions due to strain or faults in stalking of hexagonal ferrites layers, which tend to be relaxed by annealing. Similar relaxation of inner distortions by heat treatment was reported by Kaur et al. [28] for M-type hexagonal ferrite.

Table 2: Results of the refinement of the patterns of the two Y-type hexaferrite samples.

| Sample name | $\chi^2$ | Lattice Parameters (Å) | | $R_B$ | $R_F$ |
|---|---|---|---|---|---|
| | | $a = b$ | $c$ | | |
| $Co_2$-Y | 1.23 | 5.87 | 43.54 | 3.28 | 2.90 |
| $Co_2$-$Y_R$ | 1.30 | 5.86 | 43.50 | 2.63 | 2.07 |

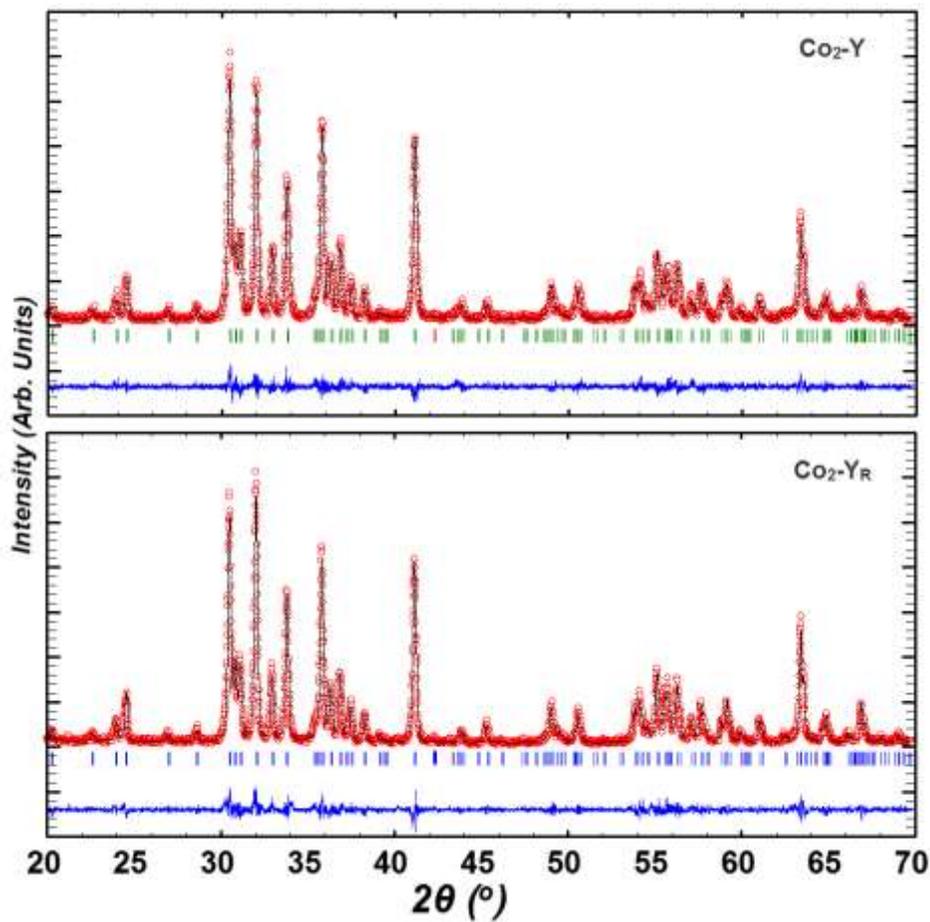

Fig. 2: Rietveld refinement results for the $Ba_2Co_2Fe_{12}O_{22}$ sample

The ionic position, and site occupancies of the samples $Co_2$-Y and $Co_2$-$Y_R$ are shown in Table 3 and 4, respectively. With the exception of the oxygen anions (O3, O4 and O5) in the 18h sites, both samples show small differences in the fractional ionic positions. The small shifts in ionic positions could be due the relaxations of inner distortions by annealing [28].

Table 3: Ionic positions and site occupancies obtained from the refinement of the $Co_2$-Y pattern.

| Atom | Site | X | Y | Z | Occupancy |
|---|---|---|---|---|---|
| Ba  | 6c       | 0.0000 | 0.0000 | 0.3002 | 1.0000 |
| Fe1 | $6c_{IV}$  | 0.0000 | 0.0000 | 0.3763 | 1.0000 |
| Fe2 | $6c_{IV}$  | 0.0000 | 0.0000 | 0.1525 | 0.6667 |
| Co1 | $6c_{IV}$  | 0.0000 | 0.0000 | 0.1525 | 0.3333 |
| Fe3 | $6c_{VI}$  | 0.0000 | 0.0000 | 0.0655 | 1.0000 |
| Fe4 | 3b       | 0.0000 | 0.0000 | 0.5000 | 0.6667 |
| Co2 | 3b       | 0.0000 | 0.0000 | 0.5000 | 0.3333 |
| Fe5 | 18h      | 0.5054 | 0.4946 | 0.1098 | 0.8889 |
| Co3 | 18h      | 0.5054 | 0.4946 | 0.1098 | 0.1111 |
| Fe6 | 3a       | 0.0000 | 0.0000 | 0.0000 | 0.6667 |
| Co4 | 3a       | 0.0000 | 0.0000 | 0.0000 | 0.3333 |
| O1  | 6c       | 0.0000 | 0.0000 | 0.4200 | 1.0000 |
| O2  | 6c       | 0.0000 | 0.0000 | 0.1986 | 1.0000 |
| O3  | 18h      | 0.1622 | 0.1622 | 0.0286 | 1.0000 |
| O4  | 18h      | 0.1819 | 0.1819 | 0.0847 | 1.0000 |
| O5  | 18h      | 0.1875 | 0.1875 | 0.1386 | 1.0000 |

Table 4: Ionic positions and site occupancies obtained from the refinement of the $Co_2$-$Y_R$ pattern.

| Atom | Site | X | Y | Z | Occupancy |
|---|---|---|---|---|---|
| Ba  | 6c       | 0.0000 | 0.0000 | 0.3001 | 1.0000 |
| Fe1 | $6c_{IV}$  | 0.0000 | 0.0000 | 0.3764 | 1.0000 |
| Fe2 | $6c_{IV}$  | 0.0000 | 0.0000 | 0.1522 | 0.6667 |
| Co1 | $6c_{IV}$  | 0.0000 | 0.0000 | 0.1522 | 0.3334 |
| Fe3 | $6c_{VI}$  | 0.0000 | 0.0000 | 0.0658 | 1.0000 |
| Fe4 | 3b       | 0.0000 | 0.0000 | 0.5000 | 0.6667 |
| Co2 | 3b       | 0.0000 | 0.0000 | 0.5000 | 0.3334 |
| Fe5 | 18h      | 0.5055 | 0.4945 | 0.1098 | 0.8889 |
| Co3 | 18h      | 0.5055 | 0.4945 | 0.1098 | 0.1111 |
| Fe6 | 3a       | 0.0000 | 0.0000 | 0.0000 | 0.6667 |
| Co4 | 3a       | 0.0000 | 0.0000 | 0.0000 | 0.3334 |
| O1  | 6c       | 0.0000 | 0.0000 | 0.4205 | 1.0000 |
| O2  | 6c       | 0.0000 | 0.0000 | 0.1977 | 1.0000 |
| O3  | 18h      | 0.1667 | 0.1667 | 0.0283 | 1.0000 |
| O4  | 18h      | 0.1801 | 0.1801 | 0.085  | 1.0000 |
| O5  | 18h      | 0.1864 | 0.1864 | 0.1384 | 1.0000 |

Table 5 shows the cation-anion distances (bond lengths) in the $Co_2$-Y and $Co_2$-$Y_R$ samples in all crystallographic sites. Generally, the bond lengths experienced small changes as a consequence of the heat treatment, where the average values of Fe3, Fe4/Co2 and Fe6/Co4 bond lengths increased, while the average values of Fe1, Fe2/Co1 and Fe5/Co3 bond lengths decreased with annealing. It is possible that heat treatment induced a change in the cations distributions over the various crystallographic sites. Since the ionic radius of the $Co^{+2}$ cations is larger than the radius of $Fe^{+3}$ [29], sites in $Co_2$-$Y_R$ sample which exhibited an increase in the average distances could be more occupied by $Co^{+2}$ compared with $Co_2$-Y sample. The opposite could have also happened in sites which revealed a decrease in distances, where more $Fe^{+3}$ cations are occupied by these sites.

Table5: Cation-Anion distances within the various crystallogrophic sites

| Distances For the $Co_2$-Y Sample (Å) | | | | Distances For the $Co_2$-$Y_R$ Sample (Å) | | | |
|---|---|---|---|---|---|---|---|
| Ba Polyhedral (6c) | | | | Ba Polyhedral (6c) | | | |
| (Ba)-(O3) | × | 3 | 3.2017 | (Ba)-(O3) | × | 3 | 3.1654 |
| (Ba)-(O3) | × | 6 | 2.9397 | (Ba)-(O3) | × | 6 | 2.9387 |
| (Ba)-(O4) | × | 3 | 2.7206 | (Ba)-(O4) | × | 3 | 2.7385 |
| Average | = | | 2.95 | Average | = | | 2.95 |
| Fe1 Tetrahedral ($6c_{IV}$) | | | | Fe1 Tetrahedral ($6c_{IV}$) | | | |
| (Fe1)-(O1) | × | 1 | 1.9039 | (Fe1)-(O1) | × | 1 | 1.9176 |
| (Fe1)-(O3) | × | 3 | 1.8467 | (Fe1)-(O3) | × | 3 | 1.8106 |
| Average | = | | 1.86 | Average | = | | 1.84 |
| Fe2/Co Tetrahedral($6c_{IV}$) | | | | Fe2/Co Tetrahedral($6c_{IV}$) | | | |
| (Fe2)-(O2) | × | 1 | 2.0071 | (Fe2)-(O2) | × | 1 | 1.9803 |
| (Fe2)-(O5) | × | 3 | 1.9992 | (Fe2)-(O5) | × | 3 | 1.9862 |
| Average | = | | 2.00 | Average | = | | 1.98 |
| Fe6/Co4 Octahedral (3a) | | | | Fe6/Co4 Octahedral (3a) | | | |
| (Fe6)-(O3): | × | 6 | 2.0657 | (Fe6)-(O3): | × | 6 | 2.0917 |

| Average | = | 2.07 | Average | = | 2.10 |

| Fe4/Co2 Octahedral (3b) | | | Fe4/Co2 Octahedral (3b) | | |
| --- | --- | --- | --- | --- | --- |
| (Fe4)-(O5) | × 6 | 1.9211 | (Fe4)-(O5) | × 6 | 1.9344 |
| Average | = | 1.92 | Average | = | 1.93 |

| Fe5/Co3 Octahedral(18h) | | | Fe5/Co3 Octahedral(18h) | | |
| --- | --- | --- | --- | --- | --- |
| (Fe5)-(O1) | × 1 | 2.018 | (Fe5)-(O1) | × 1 | 2.0056 |
| (Fe5)-(O2) | × 1 | 1.9652 | (Fe5)-(O2) | × 1 | 1.9857 |
| (Fe5)-(O4) | × 2 | 1.9307 | (Fe5)-(O4) | × 2 | 1.9259 |
| (Fe5)-(O5) | × 2 | 2.0563 | (Fe5)-(O5) | × 2 | 2.0542 |
| Average | = | 2.00 | Average | = | 1.99 |

| Fe3 Octahedral($6c_{VI}$) | | | Fe3 Octahedral($6c_{VI}$) | | |
| --- | --- | --- | --- | --- | --- |
| (Fe3)-(O3) | × 3 | 2.3014 | (Fe3)-(O3) | × 3 | 2.3513 |
| (Fe3)-(O4) | × 3 | 2.0277 | (Fe3)-(O4) | × 3 | 2.0108 |
| Average | = | 2.16 | Average | = | 2.18 |

The average inter-ionic distance in the tetrahedral around the Fe1 in $6c_{IV}$ site, is similar to the average ideal distance (1.86 Å) in tetrahedral site and close to the reported value of ~ 1.90 Å [20]. The effect of annealing reduced the distance in the average Fe1 bond length down to 1.84 Å.

The average bond length of 2.00 Å in the $6c_{IV}$ tetrahedral site around the Fe2/Co1 ions was higher than the reported value of 1.92 Å [20]. If we consider the occupancy of this site by 33% Co and 67% Fe which was adopted in the refinement, an ideal average bond length would be (1.90 Å). However, if we assume that this site is completely occupied by $Co^{+2}$, the average bond length would be ~ 1.97 Å, a value that is still slightly smaller than the average bond length in both samples ($Co_2$-Y and $Co_2$-$Y_R$). Since the cationic distribution could not account for the observed increase in bond length, this increase could be partially attributed to lattice distortions introduced by the shifts in the ionic positions of the oxygen

O2 and O5 anions within this site, which relaxed slightly with the heat treatment, reducing the average bod length down to 1.98 Å, which is close to the ideal value.

The 3b octahedral site surrounding the Fe4/Co2 cation with 33% Co was characterized by equal bond lengths of 1.92 Å and 1.93 Å for the samples $Co_2$-Y and $Co_2$-$Y_R$, respectively, which is lower than the reported value of 2.01 Å [20]. Even if we consider this site to be completely occupied by iron, the average distance would be 2.05 Å, which is still higher than the observed values. Thus, the reduction in bond length could be attributed to a compressive stress induced by the presence of the 3b site between tow octahedral (Fe3) sites (Figure 3, a).

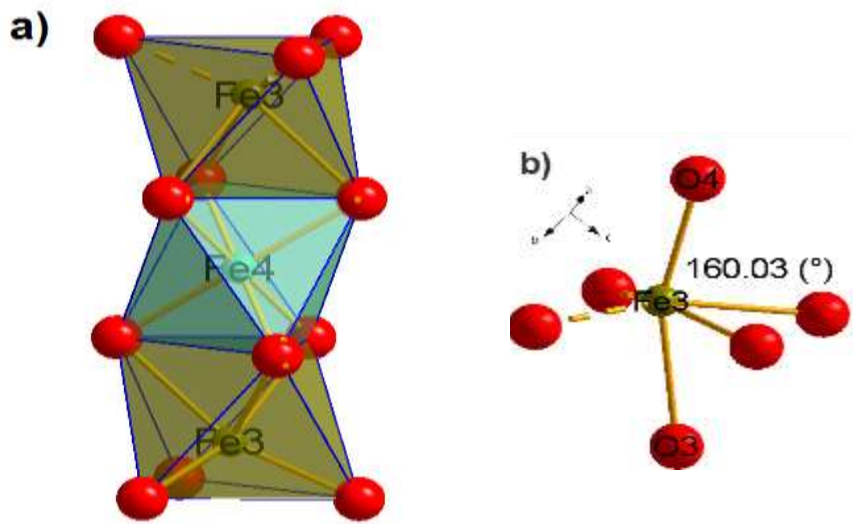

Figure 3: The configuration within the octahedral sites Fe3 and Fe4: a) face sharing octahedra and b) bond angle reduction within the octahedral around the Fe3 caused by distortion (ideal angel is supposed to be 180°)

The octahedral around the Fe3 cation in the $6c_{VI}$ site exhibits the highest distance among all Fe-O distances. This result is consistent with the results of previous studies [20, 21]. This is attributed to the face sharing between the octahedral around the Fe4 and the octahedral around the Fe3, where the octahedral around the Fe4 falls in between 2 octahedral sites around Fe3. Because of this face-sharing configuration, the Fe3-Fe4 electrostatic repulsion increases, pushing the Fe3 cation into the oxygen layer (O4), thus distorting the octahedral configuration (Figure 3, b).

The Fe6 cation in the 3a site was also found to have the same bond length with the six oxygen anions at the vertices of the octahedron. Reported value for the average cation-anion distance within this octahedral was 2.02 Å, while the occupancy revealed by Rietveld refinement of this site gave an average value of 2.07 Å. The average bond length in the sample $Co_2$-$Y_R$ increased up to 2.10 Å. This could indicate that this site is occupied by more than 33% Co. This site shares edges with other octahedra around Fe5 (18h) as shown in Figure 4.

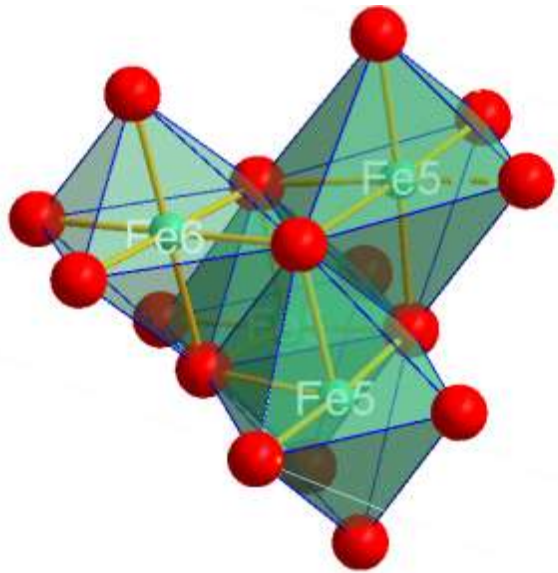

Figure 4: The configuration within the octahedrals around the Fe6 and the Fe5, where Fe6 shares some of its edges with the Fe5 configurations

The Barium-Oxygen distances are also tabulated for both samples $Co_2$-Y and $Co_2$-$Y_R$ and were found to have an average value of 2.95 Å in both samples, a value that differs slightly from previously reported values [20, 21]. The differences of bond lengths in our samples from reported values may explain the differences of the observed lattice constants (reported values: $a = 5.894$ Å, $c = 43.74$ Å).

To see a quantitative measurement of the distortion within the tetrahedral and octahedral sites, the quadratic elongation $<\Lambda_o>$ was calculated using the formula [30]:

$$<\Lambda_o> = \sum_{i=1}^{n}[(\frac{l_i}{l_o})^2/n]$$

Where $l_i$ and $l_o$ are the distances from the central cations to the vertices (anions) in the distorted and regular geometry, respectively, and $n$ is the number of distances (bonds). The results of the calculations are tabulated in table (6). It is clear from these results that the site which exhibited the largest distortion (highest deviation of the quadratic elongation from 1), was the octahedral site around Fe3/Co2 cation (Figure 3). This distortion did not seem to relax with the heat treatment, but it slightly increased possibly due to the redistribution of cobalt and iron cations in this site as a consequence of the annealing process as suggested earlier. The octahedral sites around the Fe6 and Fe5 exhibited an intermediate degree of distortion. This is attributed to the configuration they both construct (Figure 4). The remaining sites show a smaller degree of distortions in term of the quadratic elongation.

Table 6: Quadratic elongation within all crystallographic sites in the $Co_2$-Y and $Co_2$-$Y_R$

| Site | $Co_2$-Y | $Co_2$-$Y_R$ |
| --- | --- | --- |
| Fe1 | 1.0004 | 1.0021 |
| Fe2/Co1 | 1.0010 | 1.001 |
| Fe3/Co2 | 1.0350 | 1.0353 |
| Fe4/Co3 | 1.0020 | 1.0003 |
| Fe5 | 1.0137 | 1.0124 |
| Fe6/Co6 | 1.0111 | 1.0111 |

The effect of heat treatment on the micro-structural properties of the Co2-Y sample was also investigated. A standard silicon sample (Si) was used to obtain the instrumental resolution function *(U, V* and *W)* prior to performing Micro-Structural analysis. Fullprof enables the calculation of the size and strain after providing the instrumental resolution function, more details about this method is provided in the Fullprof manual [26].

For the standard Rietveld refinement the selected profile function was the Pseudo-Voigt function, while for the Micro-Structural Rietveld refinement, the selected profile function was the TCH modified Pseudo-Voigt function; this is to mimic the exact Voigt function [26, 31]. In addition, the spherical harmonics were used for the anisotropic size broadening [26, 32].

Figure (5) shows the results of the Micro-Structural Rietveld refinement. Although the refinements are completely different, no differences between the two refinements were observed (Figure 5 and 2). Table (7) shows the difference between the two refinements in terms of some parameters.

Table 7: Comparision between the Standard refinement and the Micro-Structural refinement of the $Co_2$-Y and $Co_2$-$Y_R$ Samples

| | Standard Refinement | | Micro-Structural refinement | |
|---|---|---|---|---|
| | Co2-Y | Co2-YR | Co2-Y | Co2-YR |
| $\chi^2$ | 1.23 | 1.3 | 1.23 | 1.35 |
| $R_B$ | 3.28 | 2.63 | 3.07 | 2.67 |
| $R_F$ | 2.9 | 2.07 | 2.86 | 2.26 |
| [1]$\eta$ | 0.263 | 0.4111 | - | - |
| [2]U | 0.0414 | 0.0363 | -0.0173 | 0.0004 |
| [2]V | -0.0623 | 0.0464 | - | - |
| [2]W | 0.0529 | 0.0437 | - | - |
| [3]X | 0.0108 | 0.008 | 0.0108 | 0.0106 |
| [3]Y | - | - | 0.0106 | 0.0199 |
| [4]$I_g$ | - | - | 0.0066 | 0.0014 |
| [5]$S_z$ | - | - | 0.5823 | 0.5746 |

[1] $\eta$ is the weighting parameter between the Gaussian and Lorentzian contributions to the Pseudo-Voigt function
[2] The FWHM parameters in the Caglioti formula $H^2 = U^2 \tan^2\theta + V \tan\theta + W$
[3] Refined parameters in the FWHM of the Lorentzian contribution to the TCH modified Pseudo-Voigt function
[4] Refined parameter in the FWHM of the Gaussian contribution to the TCH modified Pseudo-Voigt function equation
[5] the selected model for the Size- Anisotropic broadening (Spherical Harmonics) refined parameter

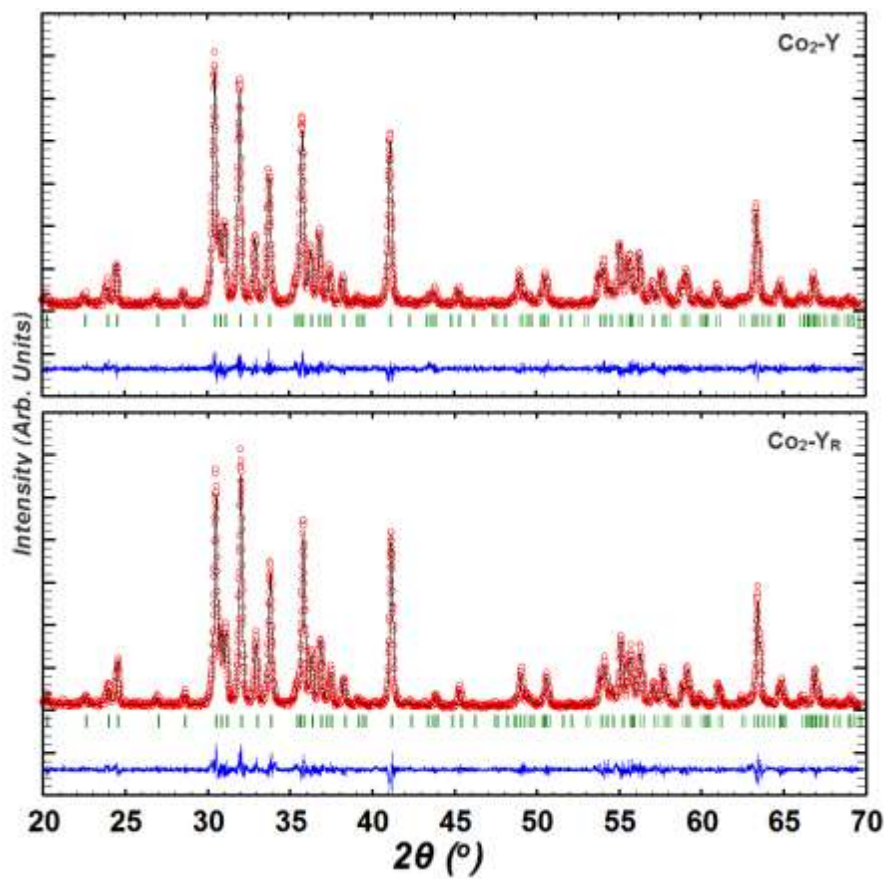

Figure 5: Micro-Structural Rietveld refinement of the $Co_2$-Y and the $Co_2$-$Y_R$ samples

Tables (8 and 9) show the calculated apparent crystallite size (size), the integral breadths of the Gaussian ($\beta_g$), and Lorentzian ($\beta_L$) contributions and the total integral breadths ($\beta$) in both samples.

Table 8: The apparent crystallite size (in unites of Angstrom Å) along some directions (hkl) and the Gaussian ($\beta_g$), Lorentzian ($\beta_l$) and total integral ($\beta$) breadths of the $Co_2$-Y sample

| h | k | l | 2theta (º) | $\beta_g$ | $\beta_l$ | $\beta$ | Size Å |
|---|---|---|---|---|---|---|---|
| | | | | (1/Å) x 1000 | | | |
| 0 | 0 | 12 | 24.5148 | 0.9207 | 0.5905 | 1.3325 | 740 |
| 1 | 1 | 0 | 30.4553 | 0.8873 | 0.8066 | 1.4676 | 660 |
| 1 | 0 | 13 | 31.9823 | 0.8775 | 0.9670 | 1.5863 | 610 |
| 1 | 1 | 6 | 32.9139 | 0.8712 | 1.0232 | 1.6261 | 600 |
| 0 | 1 | 14 | 33.7665 | 0.8652 | 0.9220 | 1.5388 | 630 |
| 1 | 1 | 9 | 35.7751 | 0.8505 | 1.1256 | 1.6910 | 570 |
| 0 | 2 | 10 | 41.1115 | 0.8065 | 1.1800 | 1.6965 | 560 |
| | | | | | | Average App-Size | 600 |

Table 9: The apparent crystallite size (in unites of Å) along some directions (hkl) and the Gaussian ($\beta g$), Lorentzian ($\beta_l$) and total integral ($\beta$) breadths of the $Co2$-$Y_R$ sample

| h | k | l | 2theta (º) | $\beta_g$ | $\beta_l$ | B | Size Å |
|---|---|---|---|---|---|---|---|
| | | | | (1/Å) x 1000 | | | |
| 0 | 0 | 12 | 24.5349 | 0.4455 | 0.9395 | 1.1789 | 880 |
| 1 | 1 | 0 | 30.4756 | 0.4472 | 0.9654 | 1.2025 | 870 |
| 1 | 0 | 13 | 32.0072 | 0.4477 | 0.9915 | 1.2251 | 860 |
| 1 | 1 | 6 | 32.9366 | 0.448 | 0.9537 | 1.1932 | 880 |
| 0 | 1 | 14 | 33.7931 | 0.4483 | 1.1098 | 1.3270 | 790 |
| 1 | 1 | 9 | 35.8007 | 0.449 | 0.9735 | 1.2109 | 870 |
| 0 | 2 | 10 | 41.1411 | 0.4511 | 1.0544 | 1.2818 | 830 |
| | | | | | | Average App-Size | 850 |

For the $Co_2$-Y sample, the value of the average crystal size is ~ 600 Å. After the heat treatment, the average value increased to have an average ~ 850 Å, which is expected, since the heat treatment increased the diffusion rate between crystallites, resulting in an increase of the crystal size.

In $Co_2$-Y sample, the crystallite size along the (0,0,12) direction was higher than that along other directions, suggesting a columnar crystallite. The heat treatment, however, seemed to remove the differences between crystallite size along different directions, indicating the effect of heat treatment in improving crystallinity of the sample in all directions.

The average maximum strain was calculated using the micro-structural Rietveld refinement. Before the heat treatment, the calculated average maximum strain was (6.61%). However, after the heat treatment the calculated value was (1.48%) suggesting that the heat treatment effectively relaxed and removed some of the distortions that were present before.

**Conclusions**

Rietveld refinement was performed using the FullProf software package. Low discrepancy values were achieved in the Rietveld refinement of the samples $Co_2$-Y and $Co_2$-$Y_R$. The heat treatment altered the general structural and micro-structural properties of the $Co_2$-Y hexagonal ferrite. Lattice constants (*a* and *c*) decreased with the heat treatment, as well as the average maximum strain, suggesting that the heat treatment process relaxed some of the distortions in the sample. The presence of the $Co^{+2}$ cations in a disordered state within the octahedral around the Fe4 increased the distortion in the Fe3 site, making it the highest distorted site in the Y-type unit cell. The distortion in the Fe3 site remained the same after the heat treatment, suggesting a higher substitution percent of $Co^{+2}$ for $Fe^{+3}$ after the heat treatment. Also, heat treatment increased the diffusion rate between crystal domains, which enhanced the formation of larger crystallites.